\documentclass[preprint,superscriptaddress,amsmath,amssymb,prd,aps,showpacs,floatfix,nofootinbib]{revtex4-1}
\usepackage{graphicx}
\usepackage{dcolumn}
\usepackage{bm}
\usepackage{mathrsfs}
\usepackage{hyperref}
\usepackage{amsmath}
\usepackage{amssymb}
\usepackage{bm}
\usepackage{color}
\usepackage{float}
\usepackage{dcolumn}
\usepackage{multirow}
\usepackage{changepage}
\usepackage{enumerate}
\hypersetup{pdftex,colorlinks=true,linkcolor=blue,citecolor=red,menucolor=black,urlcolor=blue,filecolor=blue}

\newcommand{\mev}{\textrm{ MeV}}
\newcommand{\kev}{\textrm{ keV}}

\newcommand{\eps}{\epsilon}

\newcommand{\be}{\begin{equation}}
\newcommand{\ee}{\end{equation}}
\newcommand{\ba}{\begin{eqnarray}}
\newcommand{\ea}{\end{eqnarray}}
\newcommand{\nn}{\nonumber}

\newcommand{\barb}{{\bar B}}
\newcommand{\bark}{{\bar K}}
\newcommand{\barbs}{{\bar B^*}}
\newcommand{\barks}{{\bar K^*}}
\newcommand{\bs}{B^*}
\newcommand{\ks}{K^*}

\begin{document}
\title{Exotic molecular meson states of $B^{(*)} K^{(*)}$  nature}
\date{\today}

\author{E.~Oset}
\email{oset@ific.uv.es}
\affiliation{Departamento de F\'{\i}sica Te\'orica and IFIC, Centro Mixto Universidad de
Valencia-CSIC Institutos de Investigaci\'on de Paterna, Aptdo.22085,
46071 Valencia, Spain}

\author{L. Roca}
\email[]{luisroca@um.es}
\affiliation{Departamento de F\'isica, Universidad de Murcia, E-30100 Murcia, Spain}

\begin{abstract}
We evaluate theoretically the interaction of the open bottom and strange  systems
$\barb\bark$, $\barbs \bark$, $\barb\barks$ and $\barbs\barks$ to look for possible bound states which could correspond to exotic non--quark-antiquark mesons since they would contain at least one $b$ and one $s$ quarks.
The s-wave scattering matrix is evaluated implementing unitarity by means of the Bethe-Salpeter equation, with the potential  kernels obtained from contact and vector meson exchange mechanisms. The vertices needed are supplied from Lagrangians derived from suitable extensions of the hidden gauge symmetry approach to the bottom sector. 
We find poles below the respective thresholds for isospin 0 interaction and evaluate the widths of the different obtained states by including the main sources of imaginary part, which are the $B^*\to B\gamma$ decay in the $\barbs \bark$ channels, the $K^*\to K \pi$ in the channels involving a $K^*$, plus the box diagrams with $\barb\bark$ and $\barbs\bark$ intermediate states for the $\barbs\barks$ channels.

\end{abstract}

\maketitle


\section{Introduction}

The discovery of the $X_0(2866)$ ($X_0(2900)$ officially) as a $J^P=0^+$ resonance with isospin $I=0$, decaying into $D\bar K$ \cite{LHCb:2020bls,LHCb:2020pxc} 
was an important step forward, reporting on a manifestly exotic meson state with $c$ and $s$ open quarks which, thus, cannot be accounted for as an ordinary $q\bar q$ meson. Different pictures have been proposed to explain that state as compact tetraquark structures 
\cite{Wang:2020xyc,He:2020jna,Zhang:2020oze,Wang:2020prk}.
 Yet some explicit tetraquark calculations using a relativized quark model favour instead a $D^*\barks$ molecular structure \cite{Lu:2020qmp}. The small binding of the $X_0(2900)$ with respect to the $D^*\barks$ threshold has prompted many calculations favoring the  $D^*\barks$
molecular structure \cite{Liu:2020nil,Chen:2020aos,Huang:2020ptc,Molina:2020hde,Xue:2020vtq,Agaev:2020nrc,Mutuk:2020igv,Xiao:2020ltm,He:2020btl}. Suggestions that the peaks observed could come from some kinematic singularities, as a triangle singularity, have also been done 
\cite{Liu:2020orv,Burns:2020epm}. A prediction of this state as a bound $D^*\barks$ state had already been done ten years before in 
\cite{Molina:2010tx} with results for the mass and width very similar to those reported in the experiment \cite{LHCb:2020bls,LHCb:2020pxc}. In Ref.~\cite{Molina:2020hde} a reanalysis of the work of 
\cite{Molina:2010tx} was done to fine tune the mass and width of the state and explicit decay channels were also evaluated for the companion
$D^*\barks$ states with $1^+$ and $2^+$.

The purpose of this work is to extend those results to the bottom sector studying the $\barbs\barks$ states and their decay modes. At the same time, we study the $\barb\bark$, $\barbs\bark$, and $\barb\barks$ systems and make predictions for binding energies and widths. Contrary to the 
$D^*\barks$ states that have attracted much attention, this is not the case of the $B^{(*)}K^{(*)}$ states. Yet, a study of these states using the formalism of the local hidden gauge
\cite{Bando:1984ej,Bando:1987br,Birse:1996hd,hidden4,Nagahiro:2008cv}
employed in \cite{Molina:2010tx,Molina:2020hde} is done in \cite{Kong:2021ohg}, where some exotic $\bs\ks$, $B\ks$ are found with small binding and width. We shall discuss the analogies and differences from that work, anticipating that we obtain more bound states, more binding and larger widths. Related molecular states were found in 
\cite{Sun:2018zqs} where the non exotic $\bs\barks$, $\bs\bar K$ and $B\bar K$ states were studied. Exotic states of $B^{(*)}D^{(*)}$ nature were studied in \cite{Sakai:2017avl}. The discovery of the $X_0(2900)$  and the large attention given to it, makes it opportune to study the natural extension to the bottom sector. Our strategy to make as accurate predictions as possible is to start from the results obtained in Ref.~\cite{Molina:2020hde} to fit the $X_0(2900)$ data from the $D^*\barks$ molecular perspective, using a cutoff to regularize the loops. The reason is that in the transition from the $D$ to the $B$ sector, heavy quark symmetry imposes constraints that are satisfied if one uses the same cutoff from one sector to the other 
\cite{Lu:2014ina,Altenbuchinger:2013vwa}. With these constraints we obtain several bound states of $I=0$, one of $BK$ nature with $0^+$, one $1^+$ bound state of $B^*K$ nature, another $1^+$ state of $BK^*$ nature, and three states of $B^*K^*$ nature with total $J^P=0^+,1^+,2^+$. The binding energies are bigger than for the related $D^*\barks$ systems, which seems a general trend for explicit calculations using quark models \cite{jmu,zouzou,tjon,Ke:2021rxd}.

\section{Formalism}

\subsection{Lagrangians}

\begin{figure}[h]
\centering
\includegraphics[width=.95\linewidth]{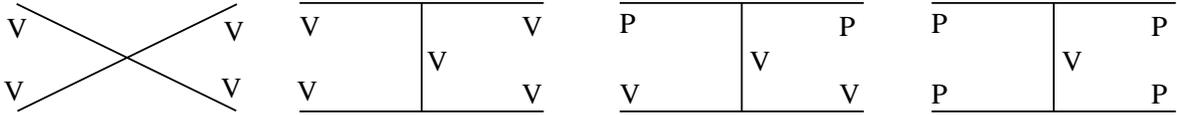}
\caption{Kinds of diagrams needed for the tree level potentials.}
\label{fig:figdiag1}
\end{figure}

The tree level scattering amplitudes needed for the different channels are of the kind depicted in Fig.~\ref{fig:figdiag1}. (We will detail below which are and why the specific diagrams for each particular channel).
Therefore we need the basic vertices, vector-pseudoscalar-pseudoscalar (VPP), three vectors (VVV) and four vectors (VVVV), which we will obtain from the extension of the local hidden gauge symmetry (HGS) formalism \cite{Bando:1984ej,Bando:1987br,Birse:1996hd,hidden4,Nagahiro:2008cv}  to the beauty sector 
\cite{Molina:2010tx,Dai:2022ulk}. The HGS formalism  has proven itself as a  suitable and successful way to realize chiral symmetry in $SU(3)$ in the presence of vector mesons, and provides the needed Lagrangians:
 \begin{eqnarray}\label{eq:VPP}
  {\cal{L}}_{VPP} &=& -ig \,\langle [P, \partial_\mu P] V^\mu\rangle \,,\\
  {\cal{L}}_{VVV} &=& ig \,\langle (V^\mu \partial_\nu V_\mu-\partial_\nu V^\mu V_\mu) V^\nu\rangle \,,\label{eq:VVV}\\
\mathcal{L}_{VVVV}&=&\frac{g^2}{2}\langle V^{\mu}V^{\nu}V^{\mu}V^{\nu}-V^{\nu}V^{\mu}V^{\mu}V^{\nu}\rangle,
\label{eq:VVVV} \end{eqnarray} 
 with $g =\frac{M_V}{2\,f}$ for which we take $M_V=800 \mev$, $f=93 \mev$. 

For the evaluation, later on, of the box diagrams arising in the calculation of the width of the $\barbs\barks$ generated state, see Fig.~\ref{fig:figbox}~b),
we will also need the Lagrangian involving the vector-vector-pseudoscalar ($VVP$) vertex, which is related to the non-abelian anomaly \cite{Wess:1971yu,witten}, and is given by 
\cite{Bramon:1992kr}:
\begin{equation}
{\cal L}_{VVP} = \frac{G'}{\sqrt{2}}\epsilon^{\mu \nu \alpha \beta}\langle
\partial_{\mu} V_{\nu} \partial_{\alpha} V_{\beta} P \rangle,
\label{eq:LVVP}
\end{equation}
with 
 $G'=\frac{3g'^2}{4\pi^2f}$,
$g'=-\frac{G_VM_{\rho}}{\sqrt{2}f^2}$ 
and $f=93\,$MeV, with $G_V=55\mev$.

In Eqs.~\eqref{eq:VPP}-\eqref{eq:LVVP}, $P$ and $V$  
are  the $q\bar{q}$ matrices which, considering only the quarks $u$, $d$, $s$ and $b$, read
\begin{eqnarray}
P=\left(
\begin{array}{cccc}
\frac{\eta}{\sqrt{3}}+\frac{\eta'}{\sqrt{6}}+\frac{\pi^0}{\sqrt{2}} & \pi^+ & K^+&B^+\\
\pi^- &\frac{\eta}{\sqrt{3}}+\frac{\eta'}{\sqrt{6}}-\frac{\pi^0}{\sqrt{2}} & K^{0}&B^0\\
K^{-} & \bar{K}^{0} &-\frac{\eta}{\sqrt{3}}+\sqrt{\frac{2}{3}}\eta'&B_s^0\\
B^-& \bar{B}^0&\bar{B}^0_s&\eta_b
\end{array}
\right)\,,
\label{eq:pfields}
\end{eqnarray}
with the standard $\eta,\eta^{\prime}$ mixing of \cite{Bramon:1992kr} and

\begin{eqnarray}
V=\left(
\begin{array}{cccc}
\frac{\omega+\rho^0}{\sqrt{2}} & \rho^+ & K^{*+}&B^{*+}\\
\rho^- &\frac{\omega-\rho^0}{\sqrt{2}} & K^{*0}&B^{*0}\\
K^{*-} & \bar{K}^{*0} &\phi&B_s^{*0}\\
B^{*-}&\bar{B}^{*0} &\bar{B}^{*0}_s&\Upsilon
\end{array}
\right)\,.
\label{eq:vfields}
\end{eqnarray}
It is worth mentioning that, as discussed in Ref.~\cite{Sakai:2017avl}, they should not be interpreted as $SU(4)$ Lagrangians, since they are only a practical way to obtain the different couplings of the vertices and is equivalent to use only $SU(3)$ considering the heavy mesons as spectators, in the line of the heavy quark flavour symmetry. Yet, in the case of meson couplings one does not even have to invoke $SU(3)$ symmetry since the couplings are tied to the simple $q\bar q$ structure of the mesons \cite{Sakai:2017avl}.

At this point we can take advantage of the work in Ref.~\cite{Molina:2010tx} regarding the $D^*\barks$ interaction. In the
$D^*\barks$ case it was only found attraction in isospin $I=0$  \cite{Molina:2010tx}, being the $I=1$ repulsive, and, as we will see, the isospin structure is equivalent in the present case, so we can expect the same result. Indeed, if we consider the $I=0$ ${D^*}\barks$ combination\footnote{The isospin doublets are $({D^*}^+,-{D^*}^0)$ and $({K^*}^0,-{K^*}^-$).}
\begin{eqnarray}\label{eq:kk1}
|D^*\barks, I=0\rangle=\frac{1}{\sqrt{2}}|{D^*}^0\bar{K}^{*0}-{D^*}^+ {K^*}^-\rangle\, ,
\end{eqnarray}
and  for $\barbs\barks$,
\begin{eqnarray}
|\barbs\barks, I=0\rangle=\frac{1}{\sqrt{2}}|{B^*}^-\bar{K}^{*0}-\bar{B}^{*0} {K^*}^-\rangle\, ,
\label{eq:kk2}
\end{eqnarray}
we see that 
Eq.~\eqref{eq:kk2} is equivalent to \eqref{eq:kk1} with the replacements ${D^*}^0\to{B^*}^-$ and ${D^*}^+\to \bar{B}^{*0}$.
This is a consequence of the fact that 
$\barbs\barks$ can be constructed from $D^*\barks$ with the only replacement of one quark $c$ by one $b$. Thus, the $q\bar q$ matrix involving $D^*$ has the same structure as Eq.~\eqref{eq:vfields} with the replacement $c\to b$:

\begin{eqnarray}
V=\left(
\begin{array}{cccc}
\frac{\omega+\rho^0}{\sqrt{2}} & \rho^+ & K^{*+}&\bar{D}^{*0}\\
\rho^- &\frac{\omega-\rho^0}{\sqrt{2}} & K^{*0}&D^{*-}\\
K^{*-} & \bar{K}^{*0} &\phi&D^{*-}_s\\
D^{*0}&D^{*+}&D^{*+}_s&J/\psi\end{array}
\right)\,.
\label{eq:vfields}
\end{eqnarray}

This is far from being a trivial detail, because this means that the tree level amplitudes arising from the Lagrangians 
\eqref{eq:VPP}-\eqref{eq:VVVV} are the same as in the ${D^*}\barks$ case with the aforementioned substitutions and changing the corresponding masses.
Thus we refer to the work in \cite{Molina:2010tx} for the specific details of the calculation of the tree level amplitudes, and we focus in what follows in the differences and peculiarities of the present work.
In particular, for the other  channels, ($\barb\bark$, $\barbs \bark$ and $\barb\barks$), which have non analogous counterparts in Ref.~\cite{Molina:2010tx}, the flavour structure is the same as in the $VV$ case and hence the numerical coefficients arising from the Lagrangians are the same except for the contribution of the different polarization vectors, which we discuss below.

\subsection{$\barb\bark$ system}

\subsubsection{Potential}

\begin{figure}[h]
\centering
\includegraphics[width=.6\linewidth]{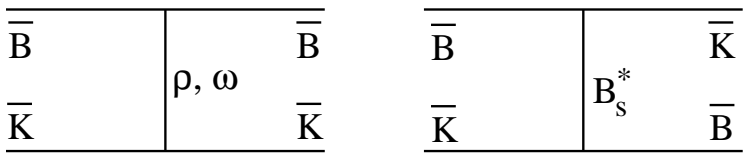}
\caption{Vector meson exchange for the $\barb\bark$ potential.}
\label{fig:figdiagBK}
\end{figure}

For the $\barb\bark$ interaction in $I=0$
the tree level potential comes from the  vector exchange mechanism depicted in Fig.~\ref{fig:figdiagBK}. From the Lagrangian \eqref{eq:VPP}, and using the appropriate charge combination of diagrams implied from Eq.~\eqref{eq:kk2}, we obtain, analogously to table XI in Ref.~\cite{Molina:2010tx}\footnote{ 
The $\barb\bark$ interaction is the same as the $\barbs\barks$ interaction due to vector exchange, except for the factor $\vec \epsilon_1\cdot\vec\epsilon_3\,\vec \epsilon_2\cdot\vec\epsilon_4$ in the latter case for $\rho$, $\omega$, exchange with $\vec\epsilon_i$ the polarization vectors of the vector mesons in the order $1+2\to 3+4$, and 
  $\vec \epsilon_1\cdot\vec\epsilon_4\,\vec \epsilon_2\cdot\vec\epsilon_3$ for the exchange of $B_s^*$. 
Since  $\vec \epsilon_1\cdot\vec\epsilon_3\,\vec \epsilon_2\cdot\vec\epsilon_4={\cal P}^{(0)}+{\cal P}^{(1)}+{\cal P}^{(2)}$  and $\vec \epsilon_1\cdot\vec\epsilon_4\,\vec \epsilon_2\cdot\vec\epsilon_3={\cal P}^{(0)}+{\cal P}^{(1)}-{\cal P}^{(2)}$, where ${\cal P}^{(J)}$ are the projector operators over the vector-vector spin $J$ states, (see Eq.~\eqref{eq:projmu}), the $\barb\bark$ interaction is the same as the one in $\barbs\barks$ for the channels of spin 0 and 2.
  },

\begin{eqnarray}
V_{\barb\bark\to\barb\bark}=-\frac{g^2}{m^2_{B_s^*}}(p_1+p_4)(p_2+p_3)
+\frac{1}{2}g^2\left(\frac{1}{m_\omega^2}-\frac{3}{m_\rho^2}
\right)(p_1+p_3)(p_2+p_4)
\label{eq:VBK}
\end{eqnarray}
with $p_1$($p_3$) the four-momentum of the initial (final) $\barb$ meson and $p_2$($p_4$) the four-momentum of the initial (final) $\bark$ meson.

We now  must project Eq.~\eqref{eq:VBK} into s-wave for which we use  that the s-wave projection of the momentum structures gives \cite{Roca:2005nm}
\begin{eqnarray}\label{eq:p1p3p2p4}
(p_1+p_3)(p_2+p_4)&\to \frac{1}{2}\big[3s-(m_1^2+m_2^2+{m_3}^2+{m_4}^2)-\frac{1}{s}(m_1^2-m_2^2)
({m_3}^2-{m_4}^2)\big]\,,\\
(p_1+p_4)(p_2+p_3)&\to \frac{1}{2}\big[3s-(m_1^2+m_2^2+{m_3}^2+{m_4}^2)+\frac{1}{s}(m_1^2-m_2^2)
({m_3}^2-{m_4}^2)\big]\,,
\label{eq:p1p4p2p3}
\end{eqnarray}
with $\sqrt{s}$ the center of mass energy and $m_i$ the mass of the particle with momentum $p_i$.
This potential at threshold takes the value $-18.7g^2$ which implies an attractive interaction and then we can expect that it  leads to a bound state after the unitarization procedure explained below.

\subsubsection{Implementation of unitarity}
\label{subsub:unit}

The full $\barb\bark$ scattering matrix can be obtained  using the techniques of the chiral unitary approach to implement unitarity building upon the elementary potential. This can be 
  carried out by
means of the Bethe-Salpeter equation, (or alternatively the $N/D$
\cite{Oller:1998zr,Oller:2000fj} or IAM
\cite{Dobado:1996ps,Oller:1998hw} methods, which are basically equivalent):
\be
T=[1-VG]^{-1}V\, ,
\label{eq:T}
\ee
where $V$ is the kernel, (the potential from Eq.~\eqref{eq:VBK}), and $G$ is the $\barb\bark$ loop function:
\begin{equation}
G=i\int\frac{d^4q}{(2\pi)^4}\frac{1}{q^2-m_B^2+i\epsilon}\frac{1}{(q-P)^2-m_K^2+i\epsilon}\ ,
\label{eq:loopex}
\end{equation}
dependent on the initial total four momentum $P$.
 The loop function $G$, since it is logarithmically divergent, needs to be properly regularized, which can be done 
using dimensional regularization, giving
 \begin{eqnarray}
G&=&\!\! \frac{1}{16 \pi^2} \left\{ a(\mu) + \ln
\frac{m_1^2}{\mu^2} + \frac{m_2^2-m_1^2 + s}{2s} \ln \frac{m_2^2}{m_1^2} 
\right. \label{propdr} +\\ \!\! &+& \!\!  
\frac{q}{\sqrt{s}}
\left[
\ln(s-(m_1^2-m_2^2)+2 q\sqrt{s})+
\ln(s+(m_1^2-m_2^2)+2 q\sqrt{s}) \right. \nonumber  - \\ 
 \!\!&-&\!\! \left. \ln(s-(m_1^2-m_2^2)-2 q\sqrt{s})-
\ln(s+(m_1^2-m_2^2)-2 q\sqrt{s}) -2\pi i \right]
\Bigg\},
\nonumber
\end{eqnarray}
where $q$ is the three-momentum of any of the intermediate particles in the center of mass frame, $\mu$ is the scale of dimensional regularization, and $m_i$ the masses of the particles in the loop.
Note that any changes
in the  scale, $\mu$,  are reabsorbed in the subtraction constant
$a(\mu)$, thus fulfilling scale invariance.
The regularization can also be carried out  implementing a three-momentum hard cutoff, $q_{\textrm{max}}$:
\begin{equation}
\label{eq:Gcutoff}
G=\int\limits_{0}^{q_{\textrm{max}}}\frac{d^3q}{(2\pi)^3}\,\frac{\omega_B+\omega_K}{2\omega_B\omega_K}\,\frac{1}{s-(\omega_B+\omega_K)^2+i\epsilon}\ ,
\end{equation}
with $\omega_{B(K)}=\sqrt{m_{B(K)}^2+\vec{q}^{\ 2}}$.
In spite of the fact that  both regularization procedures usually provide equivalent results,
it was argued in Refs.~\cite{wu,Xiao:2013jla,Ozpineci:2013qza}  
that in the heavy flavour sector the cutoff method is more appropriate since the value of the cutoff is independent of the heavy flavour, thus respecting heavy quark symmetry \cite{Lu:2014ina,Altenbuchinger:2013vwa}. 
The cutoff method is also intuitive since, using a separable potential one obtains also Eq.~\eqref{eq:T} and $q_{\textrm{max}}$ represents the range of the interaction in momentum space \cite{Gamermann:2009uq,Song:2022yvz}. This determines a natural scale for the value of the cutoff. Indeed, since the potentials are mediated by vector meson exchange, the value of the cutoff is expected to be of the order of the exchanged vector meson, $q_{\textrm{max}}\sim m_\rho$ or a bit higher since we are also exchanging a 
$B_s^*$  although  it is less relevant. In short, we have a natural scale for the cutoff, of the order of 1~GeV, but we can refine the estimation of its value using the study done in \cite{Molina:2020hde} regarding the $D^*\barks$ system. Using a similar formalism to the present work and using dimensional regularization, the subtraction constant for the $D^*\barks$ loop function  $a(\mu)$ was fitted \cite{Molina:2020hde} to reproduce the experimental mass of the $X_0(2866)$ state, obtaining $a(1500\mev)=-1.474$. We now obtain that, in order to get the same  pole position for the $D^*\barks$  interaction using the cutoff method instead, we need $q_{\textrm{max}}\simeq 1050\mev$. The same value should then be used in the present work for the bottom sector in order to preserve heavy quark mass invariance as dictated by heavy quark flavor symmetry.
Therefore, in the present work we will consider a conservative range of the cutoff, $q_{\textrm{max}}\sim 900-1050\mev$ to quantify the cutoff dependence and to get an idea of the uncertainties in our calculation.

\subsection{$\barbs\bark$ and $\barb\barks$ systems}

\subsubsection{Potential}

\begin{figure}[h]
\centering
\includegraphics[width=.6\linewidth]{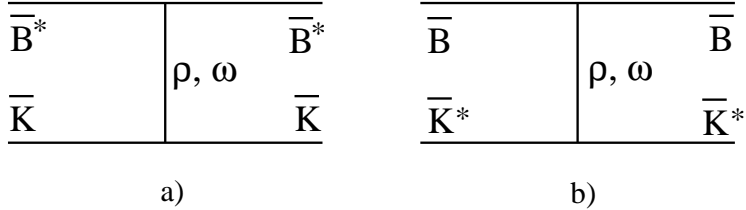}
\caption{Vector meson exchange for the $\barbs\bark$ and $\barb\barks$ potentials.}
\label{fig:figdiagBKs}
\end{figure}

The elementary diagrams for the $\barbs\bark$ and $\barb\barks$ channels, from vector meson exchange, are depicted in  Figs.~\ref{fig:figdiagBKs}a and b respectively. Note that we do not have now the $\bs_s$ exchange since it would imply an anomalous $VVP$ vertex which is small. We could also have $B_s$ exchange with ordinary $VPP$ vertices, but these crossed diagrams are largely reduced versus the vector exchange terms (see Appendix A of \cite{Dias:2021upl}) and here are further reduced with the large masses of the $B_s^*$, $B_s$. Another difference with respect to the $\barb\bark$ case is that now we have a $VVV$ vertex instead of a $VPP$ one. However, both Lagrangians (see Eqs.~\eqref{eq:VPP} and \eqref{eq:VVV}) have the same flavour structure 
and hence they provide the same potential as the $\rho,\omega$ part of Eq.~\eqref{eq:VBK} except for the vector character of the $B^*$ (or $K^*$).  This implies that, 
if we neglect terms of order $q^2/m_V^2$ \cite{Oset:2010tof},
 we have an extra $\vec \epsilon\cdot\vec \epsilon\,'$ factor in Eq.~\eqref{eq:VBK}, (where 
$\vec \epsilon(\vec \epsilon\,')$ is the polarization vector of the initial(final) vector meson):
\begin{eqnarray}
V_{\barbs\bark\to\barbs\bark}=&\frac{1}{2}g^2\left(\frac{1}{m_\omega^2}-\frac{3}{m_\rho^2}
\right)(p_1+p_3)(p_2+p_4)\,\vec \epsilon_\barbs\cdot\vec \epsilon_\barbs\,'
\label{eq:LBsK}\\
V_{\barb\barks\to\barb\barks}=&\frac{1}{2}g^2\left(\frac{1}{m_\omega^2}-\frac{3}{m_\rho^2}
\right)(p_1+p_3)(p_2+p_4)\,\vec \epsilon_\barks\cdot\vec \epsilon_\barks\,'
\label{eq:LVBKs},
\end{eqnarray}
with the substitution of Eq.~\eqref{eq:p1p3p2p4} with the masses changed accordingly. 
The values of these potentials at threshold are $-17.7g^2$ and $-31.6g^2$ for $\barbs\bark$ and $\barb\barks$ respectively which imply also a very strong attraction.
Furthermore, the polarization vectors of the vector mesons in the loop function should in principle be accounted for in the resummation implicit in the Bethe-Salpeter equation, \eqref{eq:T}. However it was shown in Ref.~\cite{Roca:2005nm} that, for the general vector-pseudoscalar interaction,  the same Bethe-Salpeter equation (\ref{eq:T}) could be used factorizing a global $\vec \epsilon\cdot\vec \epsilon\,'$, up to a correction in the loop function of $\vec q\,^2/(3m_V^2)$ which can be safely neglected.

\subsubsection{Sources of the widths}

In the $\barb\bark$ case, there is no possible source of imaginary part which could provide a width to the generated pole that we will find in the real axis below the $\barb\bark$ threshold, because both $\barb$ and $\bark$ are stable (as far as strong or electromagnetic interaction is concerned). However, in the $\barbs\bark$ and $\barb\barks$ systems the vector mesons can decay providing a finite width to the generated pole.
Indeed, for the $\barbs\bark$ system, the $B^*\to B\gamma$ decay can provide a  source of imaginary part which can give a small but finite width to the generated state. The $B^*\to B\gamma$ width has not been measured but we can take from 
\cite{cho,chengyu,jaus,slingam}
the average value 
$
\Gamma_{B^*}\simeq 0.40 \kev.$

\begin{figure}[h]
\centering
\includegraphics[width=.5\linewidth]{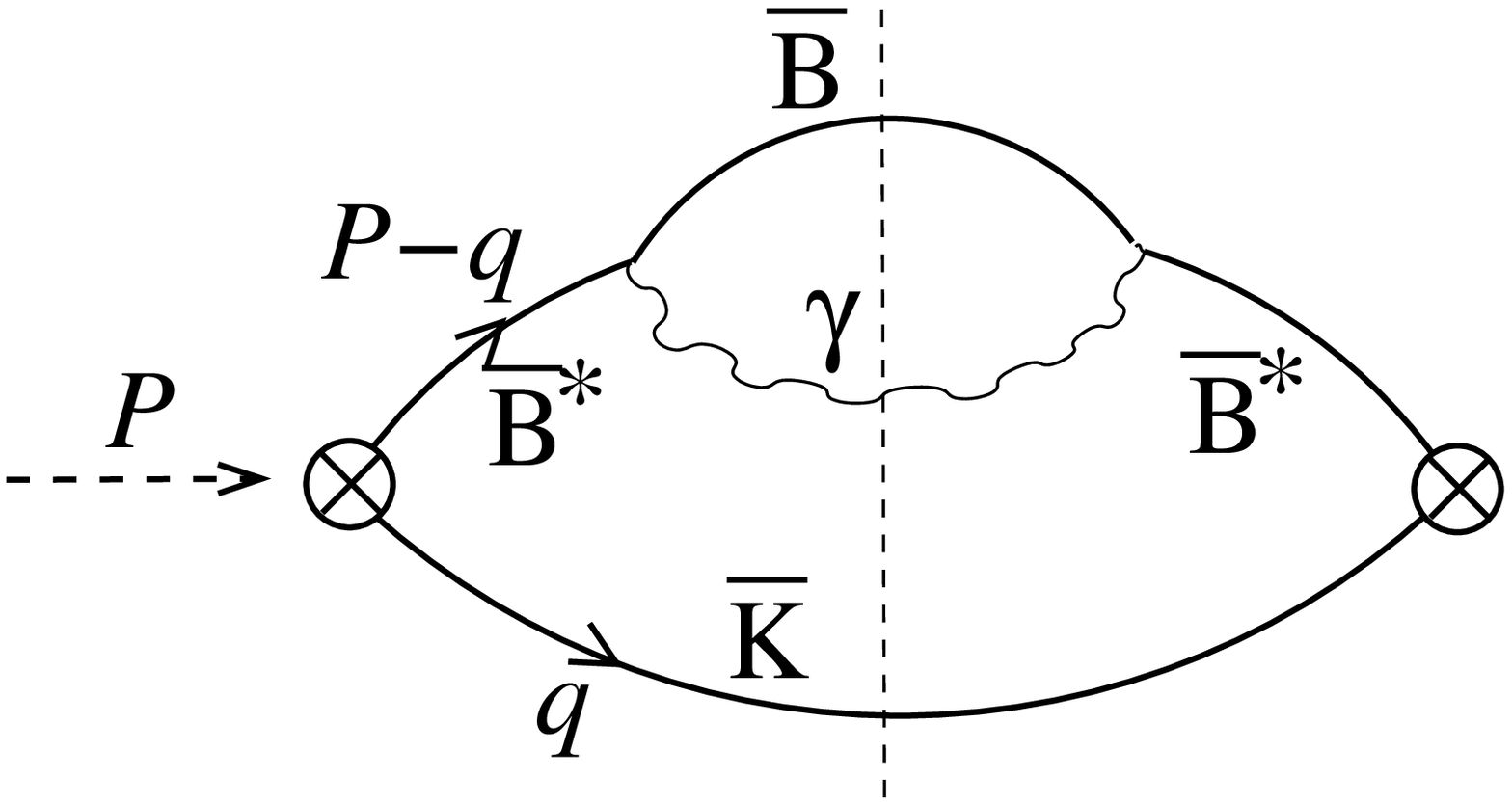}
\caption{$\barbs \bark$ loop considering the $B^*\to B\gamma$ decay. The cut shown is the source
of the imaginary part of the unitarized $\barbs \bark$ scattering amplitude.}
\label{fig:figloopBsgamma}
\end{figure}

In Fig.~\ref{fig:figloopBsgamma}   we show the mechanism including the $B^*\to B\gamma$ decay providing an imaginary part from the cut depicted in the figure. This is the source of the width of the $\barbs\bark$ generated state, which should then be of electromagnetic order of magnitude. 

This effect can be taken into account, in a similar way as in Ref.~\cite{Dai:2022ulk}, introducing the energy dependent $B^*\to B\gamma$ width into the $\barbs$ propagator  of the loop function:
\ba
G(s)=i\int\frac{d^4 q}{(2\pi)^4}\,\frac{1}{q^2-m_K^2+i\epsilon}
\,\frac{1}{(P-q)^2-m_{B^*}^2+i \sqrt{(P-q)^2}\Gamma_{B^*}((P-q)^2)}\,,
\label{eq:GwidthBs}
\ea
This way to introduce the $B^*$ width is formally more rigorous than the  usual convolution with the $B^*$ spectral function since it accounts properly for the off-shell--ness of the $B^*$ in the loop. The energy dependent 
$B^*\to B\gamma$ width is evaluated as:
\ba
\Gamma_{B^*}(s')=\Gamma_{B^*}(m_{B^*}^2) \frac{m_{B^*}^2}{s'}
\left(\frac{p_\gamma(s')}{p_\gamma(m_{B^*}^2)}\right)^3
\Theta(\sqrt{s'}-m_B)\,,
\label{eq:widthofs}
\ea
where $\Gamma_{B^*}(m^2_{B^*})\simeq 0.4\kev$ is the aforementioned on-shell width; $p_\gamma$ is the photon decay momentum and $\Theta$ is the step function.

The $q^0$ integration in Eq.~\eqref{eq:GwidthBs} leads to
\ba
G(s)\simeq \int_0^{q_{\rm max}}dq\frac{q^2}{4\pi^2}\,\frac{\omega_{B^*}+\omega_K}{\omega_{B^*}\omega_K}
\,\frac{1}{\sqrt{s}+\omega_{B^*}+\omega_K}\,
\frac{1}{\sqrt{s}-\omega_{B^*}-\omega_K+i\frac{\sqrt{s'}}{2\omega_{B^*}}
\Gamma_{B^*}(s')
}\,,
\label{eq:GwidthB}
\ea
with $\omega_{K(B^*)}=\sqrt{\vec q\,^2+m_{K(B^*)}^2}$
and $s'=(\sqrt{s}-\omega_K)^2-\vec q\,^2$.

Note that now Eq.~\eqref{eq:GwidthB} provides a small but finite imaginary part for the $T$ matrix corresponding to the cut depicted in Fig.~\ref{fig:figloopBsgamma}.

\begin{figure}[h]
\centering
\includegraphics[width=.5\linewidth]{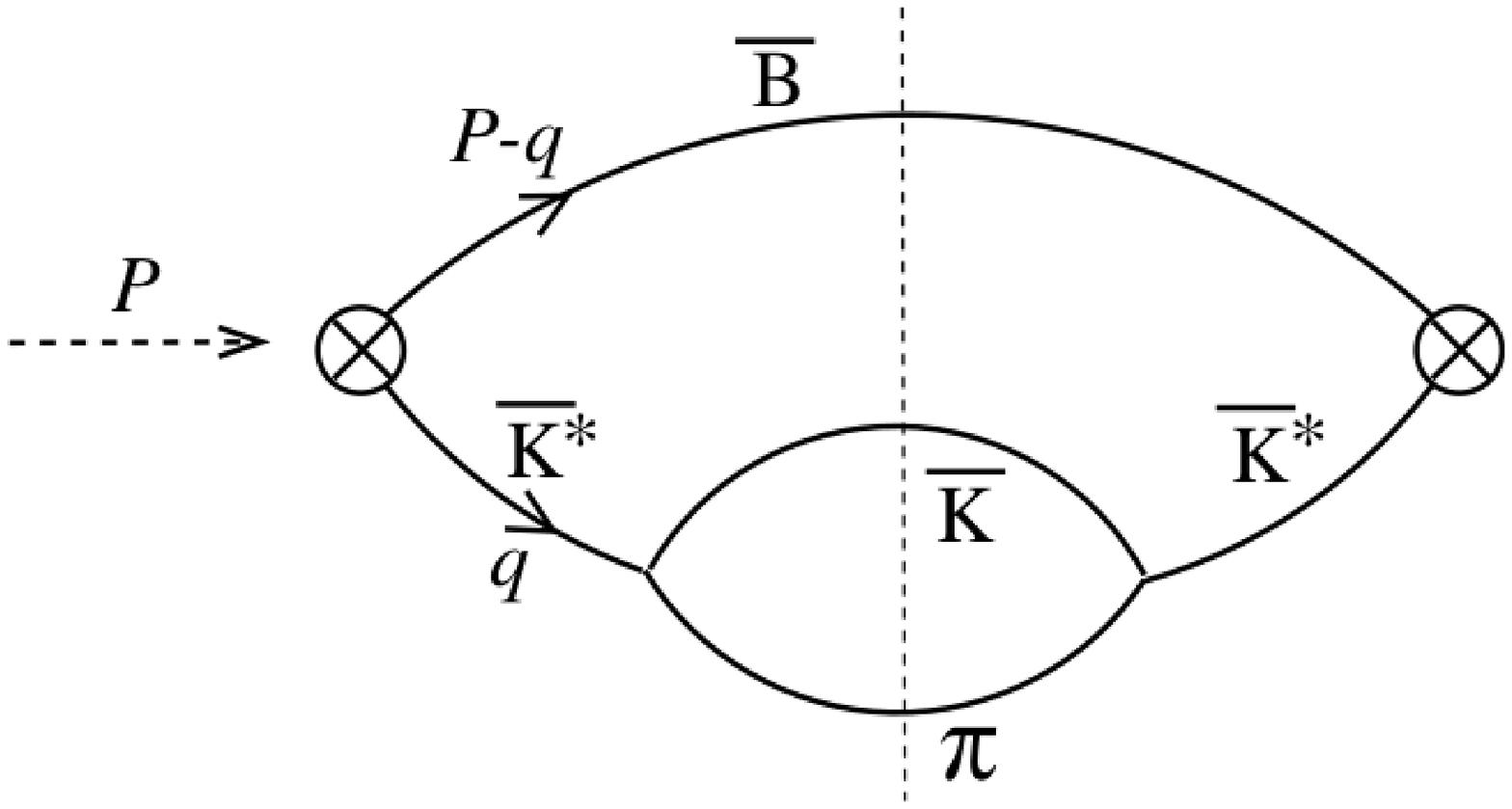}
\caption{$\barb \barks$ loop considering the $K^*\to K\pi$ decay. The cut shown is the source
of the imaginary part of the unitarized $\barb \barks$ scattering amplitude.}
\label{fig:figloopBKpi}
\end{figure}

For the $\barb\barks$ system the imaginary part comes from the cut shown in Fig.~\ref{fig:figloopBKpi}, that is, from the $K^*$ decay width. In this case the width  brought to the $\barb\barks$ generated state would be in the scale of the strong interaction and can be accounted for introducing the $K^*\to K\pi$ decay width into the loop function in an analogous way to Eq.~\eqref{eq:GwidthB}:
\ba
G(s)\simeq \int_0^{q_{\rm max}}dq\frac{q^2}{4\pi^2}\,\frac{\omega_{B}+\omega_{K^*}}{\omega_{B}\omega_{K^*}}
\,\frac{1}{\sqrt{s}+\omega_{B}+\omega_{K^*}}\,
\frac{1}{\sqrt{s}-\omega_{K^*}-\omega_B+i\frac{\sqrt{s'}}{2\omega_{K^*}}
\Gamma_{K^*}(s')
}\,,
\label{eq:GwidthK}
\ea
with $s'=(\sqrt{s}-\omega_B)^2-\vec q\,^2$,
 and
\ba
\Gamma_{K^*}(s')=\Gamma_{K^*}(m_{K^*}^2) \frac{m_{K^*}^2}{s'}
\left(\frac{p_\pi(s')}{p_\pi(m_{K^*}^2)}\right)^3
\Theta(\sqrt{s'}-m_K-m_\pi)\,,
\label{eq:widthofsKs}
\ea
with $\Gamma_{K^*}(m_{K^*}^2)$ the on-shell $K^*$ width.

\subsection{$\barbs\barks$  system}

\subsubsection{Potential}

\begin{figure}[h]
\centering
\includegraphics[width=.9\linewidth]{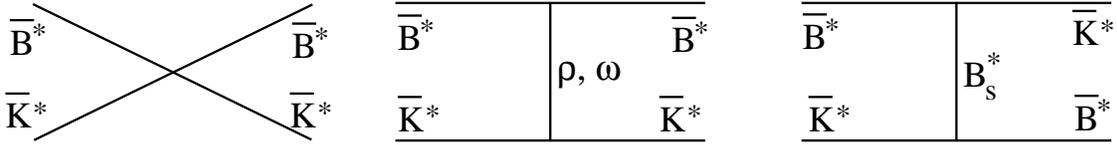}
\caption{Contact term and vector meson exchange for the $\barbs\barks$ potential.}
\label{fig:figdiagBsKs}
\end{figure}

Within the chiral unitary approach, the vector-vector interaction was studied in detail in \cite{diana,gengvec}. It was applied to $D^* D^*$, $D_s^* D^*$, $D^* K^*$ and $D^* \bar K^*$  system in \cite{Molina:2010tx}; and to $B^* B^*$ in \cite{Dai:2022ulk}.
In this case the potential arises from the direct $VVVV$ contact term plus the vector exchange contributions, depicted in Fig.~\ref{fig:figdiagBsKs}.
An s-channel vector exchange  is also possible  but it was shown in Refs.~\cite{diana,gengvec} to be very small.

The contact term gives the potential
\ba
V_{\barbs\barks\to\barbs\barks}^{\textrm{contact}}=g^2\left(
2\epsilon_\mu\epsilon^\mu \epsilon_\nu\epsilon^\nu
-\epsilon_\mu \epsilon_\nu\epsilon^\mu\epsilon^\nu
-\epsilon_\mu \epsilon_\nu\epsilon^\nu\epsilon^\mu\right)
\label{eq:V4V}
\ea
where the order of the polarization vectors represents initial $\barbs$, initial $\barks$, final $\barbs$ and final $\barks$ respectively.
Using the spin projector operators, discussed in \cite{diana},
 \begin{eqnarray}
{\cal P}^{(0)}&=& \frac{1}{3}\eps_\mu \eps^\mu \eps_\nu \eps^\nu\nonumber\\
{\cal P}^{(1)}&=&\frac{1}{2}(\eps_\mu\eps_\nu\eps^\mu\eps^\nu-\eps_\mu\eps_\nu\eps^\nu\eps^\mu)\nonumber\\
{\cal P}^{(2)}&=&\frac{1}{2}(\eps_\mu\eps_\nu\eps^\mu\eps^\nu+\eps_\mu\eps_\nu\eps^\nu\eps^\mu)-\frac{1}{3}\eps_\mu\eps^\mu\epsilon_\nu\epsilon^\nu\ ,
\label{eq:projmu}
\end{eqnarray}
for spin $J=0$, $1$ and $2$ respectively, the potential in Eq.~\eqref{eq:V4V} contributes as
\be
V_\textrm{contact}= \left\{\begin{array}{cc}\vspace{0.3cm}
4 g^2&\qquad\textrm{for $J=0$},\\
0  &\qquad\textrm{for $J=1$},\\
-2 g^2&\qquad\textrm{for $J=2$}.
\end{array}\right.
\label{eq:tcontact}
\ee

For the vector meson exchange diagrams in Fig.~\ref{fig:figdiagBsKs}, the contribution is the same as Eq.~\eqref{eq:VBK} but substituting the momentum structures by
\ba
(p_1+p_4)(p_2+p_3)\to (p_1+p_4)(p_2+p_3) \epsilon_1\cdot \epsilon_4\,\epsilon_2\cdot \epsilon_3, \nn\\
(p_1+p_3)(p_2+p_4)\to (p_1+p_4)(p_2+p_3) \epsilon_1\cdot \epsilon_3\,\epsilon_2\cdot \epsilon_4,
\ea
which, using the spin projectors \eqref{eq:projmu}, give the following contributions to the potential:

\begin{eqnarray}
V_\textrm{exch.} = \left\{\begin{array}{ccl}\vspace{0.3cm}
-&\frac{g^2}{m^2_{B_s^*}}(p_1+p_4)(p_2+p_3)
+\frac{1}{2}g^2\left(\frac{1}{m_\omega^2}-\frac{3}{m_\rho^2}
\right)(p_1+p_3)(p_2+p_4), &\textrm{ for $J=0$, 2}\\
&\frac{g^2}{m^2_{B_s^*}}(p_1+p_4)(p_2+p_3)
+\frac{1}{2}g^2\left(\frac{1}{m_\omega^2}-\frac{3}{m_\rho^2}
\right)(p_1+p_3)(p_2+p_4), &\textrm{ for $J=1$}\end{array}\right.
\label{eq:VBsKsJ}
\end{eqnarray}
where we also have to carry out the analogous s-wave projection to Eqs.~\eqref{eq:p1p3p2p4} and \eqref{eq:p1p4p2p3}, changing the corresponding masses. 
This potential at threshold takes the values $-29.2g^2$, 
$-30.6 g^2$ and $-35.2g^2$ for $J=0$, 1 and 2 respectively, which imply very strong attractive interactions for all the spins. Note that the spin degeneracy is slightly broken by the small contact and $B_s^*$ exchange terms. Note also that the exchange of light vectors, $\rho$ and $\omega$, from Eq.~\eqref{eq:VBsKsJ} goes as $V_\textrm{exch.} \sim
2m_{\bs}2m_{\ks}$, with the normalization used in our approach, which is a typical heavy quark spin symmetry behavior since the heavy quark is a spectator in the interaction which is then independent of this heavy quark (see details in \cite{Liang:2014eba}).

\subsubsection{Sources of the widths}

The first source of imaginary part that we must consider comes from the $\barbs\barks$ loop function, since we must also include the $\barks$ width in an analogous way to Eq.~\eqref{eq:GwidthK}. For this channel, the $\barbs$ decay width, being electromagnetic, is totally negligible compared to the $\barks$ one.

\begin{figure}[h]
\centering
\includegraphics[width=.7\linewidth]{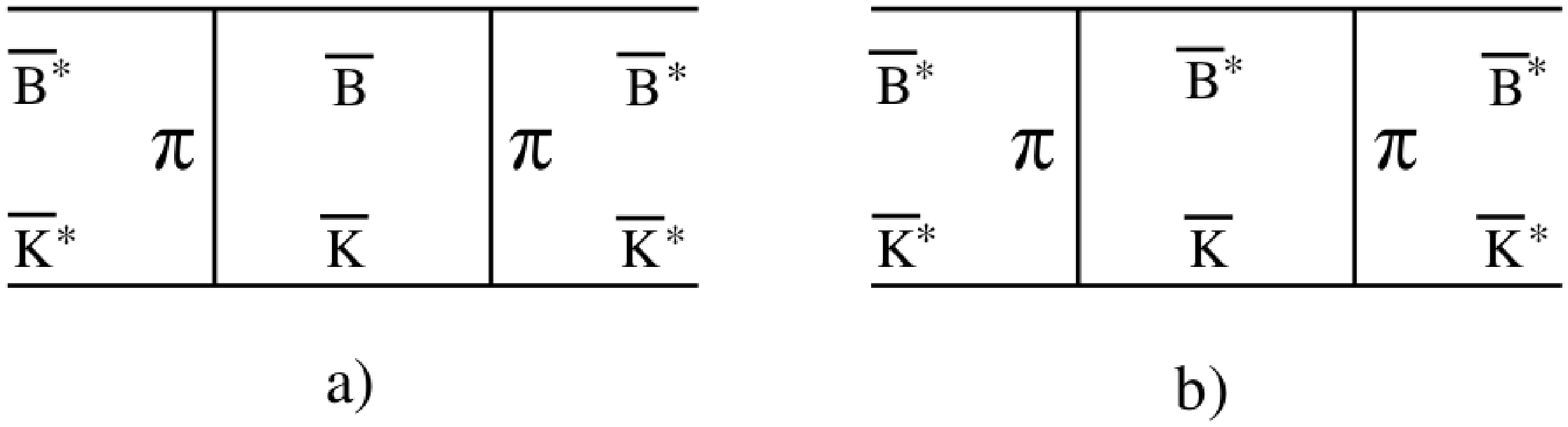}
\caption{Box diagrams for the evaluation of the width of the $\barbs\barks$ states.}
\label{fig:figbox}
\end{figure}

But now we also have the contribution from the box diagrams depicted in Fig.~\ref{fig:figbox}.
The diagram in Fig.~\ref{fig:figbox}~a), which we will call box A in the following, only gives contribution for $J=0$ and $J=2$ since the intermediate $\bar B$ and $\bar K$ need to be in $L=0$ or $2$ to match the parity and angular momentum of the initial $\barbs\barks$ state.
Since the real part of the box is expected to be small, as it is the case in $D^*\barks$ \cite{Molina:2010tx}, and we are interested in knowing its contribution to the $\barbs\barks$ state, we can focus on evaluating only the imaginary part of box A. Furthermore, since we will work  close to the $\barbs\barks$ threshold, we can safely neglect all the external three-momenta in the evaluation of the loops. With this in mind the amplitude of the box A takes the form
\ba
V_{\textrm{boxA}}=i 36 g^4\int\frac{d^4q}{(2\pi)^4}
\frac{\vec \epsilon_1\cdot\vec q\, \vec \epsilon_2\cdot\vec q
\,\vec \epsilon_3\cdot\vec q\, \vec \epsilon_4\cdot\vec q}{[q^2-m_\pi^2]^2[(p_1-q)^2-m_B^2+i\epsilon][(p_2+q)^2-m_K^2+i\epsilon]}
\label{eq:vboxadq4}
\ea
where the numerical factor in front of the integral comes from the combination of the different channels needed to have $I=0$, see Eq.~\eqref{eq:kk2}.
And, after performing the $q^0$ integration, it yields:
\ba
V_{\textrm{boxA}}=9 g^4\int\frac{d^3q}{(2\pi)^3}\frac{1}{\omega_K\omega_B}
\frac{\vec \epsilon_1\cdot\vec q\, \vec \epsilon_2\cdot\vec q
\,\vec \epsilon_3\cdot\vec q\, \vec \epsilon_4\cdot\vec q}
{[(\omega_K-p_2^0)^2-\vec q^2-m_\pi^2]^2[p_1^0+p_2^0-\omega_K-\omega_B+i\epsilon]}\, .
\label{eq:Imvboxa}
\ea
Following an analogous procedure to Ref.~\cite{Dai:2021vgf} for the $D^*D^*$ case to deal with the polarization vectors and the momentum structure of the numerator, we obtain the following imaginary part:
\ba
\textrm{Im}\, V_{\textrm{boxA}}=-3 g^4 \frac{1}{10\pi\sqrt s} q^5 
\frac{1}{ [(\omega_K-p_2^0)^2-\vec q^2-m_\pi^2]^2} F_J F^4(q) 
\left(\frac{m_{\bs}}{m_{\ks}}\right)^2
\label{eq:imta}
\ea
where $F_J$ is 5 for spin $J=0$, 0 for $J=1$ and 2 for $J=2$;
  $q=\frac{\lambda^{1/2}(s,m^2_B,m^2_K)}{2\sqrt{s}}$ and $p_2^0=\frac{s+m^2_{K^{*}}-m^2_{\bs}}{2\sqrt{s}}$.
  In Eq.~\eqref{eq:imta} we have included a factor $\frac{m_{\bs}}{m_{\ks}}$ for each $\barbs\bar B\pi$ vertex coming from the normalization of the external heavy vector meson, for the reasons explained in detail in 
\cite{Liang:2014eba}.
We have also added for each vertex, as in \cite{Molina:2010tx}, a form factor $F(q)=\exp[-q^2/\Lambda^2]$, with $q$ the four-momentum of the pion in the loop, in our case  $F(q)=\exp[(\omega_K-p_2^0)^2-\vec q\,^2]/\Lambda^2]$, with $\Lambda\sim 1200\mev$ \cite{Navarra:2001ju} to take into account the off-shell--ness of the pion in the loop obtained by using QCD sum rules \cite{Navarra:2001ju}.
Finally $i\textrm{Im}\, V_{\textrm{boxA}}$ has to be added to the contact and vector exchange contributions, (Eqs.~\eqref{eq:tcontact} and \eqref{eq:VBsKsJ}) to get the kernel, $V$, of the Bethe-Salpeter equation \eqref{eq:T}.

The contribution to the imaginary part of the potential coming from  the diagram of Fig.~\ref{fig:figbox}~b) was calculated for the $D^*\barks$ interaction in \cite{Molina:2020hde} and the present case gives an analogous result changing the corresponding masses:

\begin{eqnarray}
Im V_{\rm box B}&=&-\frac{1}{8\pi\sqrt{s}} \,q^5 
\,\left(G'g m_{B^*}\right)^2 
  \frac{1}{\left((\omega_K-p^0_2)^2-\vec q^2-m^2_{\pi}\right)^2} \,F'_J F^4(q) \, ,
\label{eq:boxB}
\end{eqnarray}
with  $F'_J=0$, $\frac{3}{2}$ and     $\frac{9}{10}$ for $J=0$, 1 and 2 respectively; and $
q=\frac{\lambda^{1/2}(s,m^2_{B^*},m^2_K)}{2\sqrt{s}}\,;\quad  p_2^0=\frac{s+m^2_{K^*}-m^2_{B^{*}}}{2\sqrt{s}}$.
Note that now the $m^2_{B^*}$ factor of heavy quark spin symmetry is already implemented by the anomalous vertex \cite{Liang:2014eba}.

The final expression for the potential, $V$, entering the Bethe-Salpeter equation \eqref{eq:T}, for the $\barbs\barks$ case, is

\be
V=V_\textrm{contact}+V_\textrm{exch.}+ i \textrm{Im}\, V_{\textrm{boxA}}
+i \textrm{Im}\, V_{\textrm{boxB}}
\ee
from Eqs.~\eqref{eq:tcontact}, \eqref{eq:VBsKsJ}, \eqref{eq:imta} and \eqref{eq:boxB}.
When evaluating the full scattering amplitude, Eq.~\eqref{eq:T}, for the $\barbs\barks$ case, we also include in the  $\barbs\barks$ loop function, $G$, the $\ks$ width in an analogous way to  Eq.~\eqref{eq:GwidthK}.


\section{Results}

\begin{table}[h]\centering\small%
\caption {Results for the binding energy, $E_B$, coupling, $g_R$, and the width without and with the consideration of the box diagrams of Fig.~\ref{fig:figbox}. 
The first number in the numerical cells represent the value obtained with the  cutoff $q_\textrm{max}=900\mev$ and the second one using $q_\textrm{max}=1050\mev$.
The number in brackets besides the name of the channel shows the value of the threshold of the corresponding channel in MeV. }
\begin{tabular}{|c|c|c|c|c|c|}%
\hline
channel    & $I(J^P)$ &$E_B$ (MeV) & $g_R$ (GeV) &\begin{tabular}{@{}c@{}} width without box \\ diagrams (MeV) \end{tabular} & full width (MeV)\\ \hline
$\barb\bark$(5774.7)  & $0(0^+)$ & 7--22 & 17--22  & -- & --	 \\ \hline
$\barbs\bark$(5820.4) & $0(1^+)$ & 3--15& 14--20  &(117-10)\textrm{ eV}& (117-10)\textrm{ eV} \\ \hline
$\barb\barks$(6172.9) & $0(1^+)$ & 70--117  & 34--38 & 6.5--1.9 & 6.5--1.9 \\ \hline
\multirow{3}{*}{$\barbs\barks$(6218.6)} &   
                $0(0^+)$ & 54-94  & 31--35	 & 9.0--3.1 & 115--160\\ \cline{2-6}
               & $0(1^+)$ & 62--106  & 33--37	 & 7.6--2.4 & 13--10 \\ \cline{2-6}
               & $0(2^+)$ & 90--145  &38--42	 & 4.2--0.9& 55--80\\
\hline
\end{tabular}
\label{tab:results1}
\end{table}

First we show in the third column of table~\ref{tab:results1} the binding energies, $E_B$, for the different channels defined as the difference between the threshold of the channel and the position of the 
pole of the scattering amplitude. (The value of the thresholds are shown in brackets besides the name of the channels). The values of $E_B$ are obtained without including the sources of imaginary part,  that is, using only the tree level potentials, (the inclusion of the imaginary parts has a small effect in  the position of the peaks).
We show the results for two different values of the cutoff, $q_\textrm{max}$, as explained in section \ref{subsub:unit}. In every numerical entry of the table the first number is the result for $q_\textrm{max}=900\mev$ and the second one for $q_\textrm{max}=1050\mev$.
 The value of the cutoff is by far the main source of error.
 Nevertheless, for the reasons explained at the end of section \ref{subsub:unit}, the results for $q_\textrm{max}=1050\mev$ should be considered as more reliable and the other value of the cutoff can be understood as a measure on the cutoff dependence and, to a lesser extent, as an estimation of the uncertainty in our calculations.
  
The column labeled ``width without box diagrams'' shows the values of the width of the generated states including in the loop functions the source of imaginary part from only Fig.~\ref{fig:figloopBsgamma} and \ref{fig:figloopBKpi}, that is, without the box mechanisms in Fig.~\ref{fig:figbox}.
Note that the width for the $\barbs\bark$ channel is very small, on the electromagnetic scale, as a consequence of  the estimated  $0.4 \kev$ of the $B^* \to B \gamma$ decay width.
 In the last column we show the final result of the width, that is, including also the imaginary source from Fig.~\ref{fig:figbox}, in addition to Fig.~\ref{fig:figloopBsgamma} and \ref{fig:figloopBKpi}. The widths are obtained directly from the  plots of $|T|^2$.
In the  fourth column we also show the value of the 
couplings of the different states to the corresponding meson-meson channel, which is obtained as the residue of the scattering amplitude, $T$, at the pole position since, if the pole is close to the real axis, we can define
\begin{eqnarray}
T\simeq\frac{g_R^2}{s-s_R},
\end{eqnarray}
with $s_R\equiv M_R^2$ the squared energy of the bound state.
Therefore
\ba
g_R^2=\lim_{s \to s_R} (s-s_R)\, T(s),
\ea
which is just the residue at the pole.
The values of the couplings shown in table \ref{tab:results1} are  for the case without including the imaginary parts, since this effect has a small influence in the couplings.

We can see in the table that we find typically large values of the binding energies. For the $\barbs\barks$ system the values 94, 106 and 145$\mev$ for J=0, 1 and 2 respectively,  are larger than their analogous ones for $D^*\barks$ \cite{Molina:2020hde}, which are 38, 43 and $129\mev$ respectively.
This increase of the binding energy with the mass of the heavy meson
is a common observation in other works comparing the charm and bottom sectors \cite{jmu,zouzou,tjon,Ke:2021rxd,Dai:2022ulk}.

At this point it is opportune to compare the results obtained with those of Ref.~\cite{Kong:2021ohg}. This latter work does not aim at making precise predictions for the $\barb^{(*)}\bark^{(*)}$ states. A global form factor is used that suppresses the propagators when the particles are off-shell, the $q^0$ integration of the loops is also done in a different way invoking the covariant spectator approximation of Ref.~\cite{Gross:1991pm} and results are shown  for different values of the $\Lambda$ parameter of the form factor. Yet, in the range of values of $\Lambda$ chosen, the binding energies obtained in \cite{Kong:2021ohg} are smaller than ours by at least one order of magnitude. The widths obtained there are also much smaller than those found in the present work. We share some qualitative features, however. Indeed, only $I=0$ states are found, like in the present approach. The $1^+$ $\bs K$ states are not bound in \cite{Kong:2021ohg} for the range of values of $\Lambda$ chosen, while bindings up to 24~MeV are found for the $B\ks$ system in the same range of $\Lambda$. In table~\ref{tab:results1} we see that we get binding for both systems, but the binding energies of $\barb\barks$ are about one order of magnitude bigger than for $\barbs\bark$.

On the other hand, in Ref.~\cite{Kong:2021ohg} the authors also calculate  the binding for the $\bar D^* \ks$ state, and for $0^+$ they find bindings of the order of $1-5\mev$ in the range of $\Lambda$ values considered. Yet, the actual $0^+$ state $X_0(2866)$ considered as a  $\bar D^* \ks$ molecule is bound by about $34\mev$. On the other hand, going from $0^+$ $\bar D^* \ks$ system  to the $\bs\ks$ one, the binding energies do not change much in \cite{Kong:2021ohg}, while in our case they are increased by about a factor of three, in agreement with general arguments found in different works  \cite{Dai:2022ulk,jmu,zouzou,tjon,Ke:2021rxd} regarding the scaling with the mass of the heavy meson. As to the small widths of \cite{Kong:2021ohg}, obtained by using coupled channels, the factor $({m_{\bs}}/{m_{\ks}})^2$ of Eq.~\eqref{eq:imta} for the A mechanism of decay, demanded to obtain a coupling  of $\bs$ to $B\pi$ consistent with lattice results \cite{Flynn:2013kwa}
and heavy quark symmetry \cite{Liang:2014eba}, is partly responsible for the large widths obtained in the present work and is apparently missing in 
\cite{Kong:2021ohg}. This could partly  explain the large differences in the widths.  The prospective work of \cite{Kong:2021ohg} has its value showing that states are only obtained in $I=0$ and providing an idea of where one can observe new states. The present work, relating the $D^{(*)}\bark^{(*)}$ and $\barb^{(*)}\bark^{(*)}$ systems through arguments of heavy quark symmetry \cite{Lu:2014ina,Altenbuchinger:2013vwa} and using the experimental values for the mass and width of the $X_0(2900)$ should provide accurate results which should encourage the experimental search for the different states observed.

\section{Conclusions}

We have studied the interaction $\barb\bark$, $\barbs \bark$, $\barb\barks$ and $\barbs\barks$ in s-wave  and isospin 0 using the techniques of the chiral unitary approach to resum the multiple final state interaction  implied in the unitarization procedure through the Bethe -Salpeter equation. The basic potentials are based on the dominant vector exchange interaction (plus contact terms in the $\barbs\barks$ case), obtained from suitable Lagrangians from the local hidden gauge symmetry formalism to deal with vector mesons, extended to the bottom sector in a way successfully tested  in many previous works.
The main source of uncertainty in the present work is the value of the regulator of the logarithmically divergent meson-meson loop function entering the unitarization formalism, accounted for by means of a three-momentum cutoff, which is obtained from reproducing the experimental pole position of the $X_0(2900)$ state in the $\bar D^* \ks$ interaction.
 We find poles for all the $\barb^{(*)} \bark^{(*)}$ channels and for all the range of reasonable values of the cutoff considered. We also evaluate the widths of the generated states through the inclusion of the main sources of imaginary part, that is, the width of the $\barks$ (or the electromagnetic $\barbs$ decay in the $\barbs\bark$ case since this is the only possible source) and the box diagrams containing  $\barb\bark$ and $\barbs\bark$ intermediate mechanisms for the $\barbs\barks$ systems.

 Despite the uncertainty obtained, it is a grounded and sound conclusion of the present study that these exotic states must exist and thus it should spur experimental efforts to try to find them.

\section*{ACKNOWLEDGEMENT}
 This work is 
supported by the Spanish Ministerio de Economia y Competitividad and European FEDER funds under
Contracts No. FIS2017-84038-C2-1-P B and by Generalitat Valenciana under contract No. PROMETEO/
2020/023. This project has received funding from the European Unions Horizon 2020 research and innovation
programme under grant agreement No.824093 for the STRONG-2020 project.
ely.


\end{document}